\documentclass[11pt]{amsart}
\usepackage{fullpage}
\usepackage{amsmath, amssymb, amsthm,mathrsfs}
\usepackage{enumerate}
\usepackage{algorithm}
\usepackage {algpseudocode}
\usepackage{graphicx}
\usepackage{color}
\usepackage{bm}
\usepackage{bigints}
\usepackage{textgreek}
\usepackage[colorlinks=true,linkcolor=red,citecolor=blue,urlcolor=cyan]{hyperref}
\usepackage{verbatim}
\usepackage{enumitem}
\usepackage{mathtools}
\usepackage{longtable}

\theoremstyle{remark}

\usepackage[backend=bibtex,giveninits=true,sorting=nyt,natbib=true,maxcitenames=2,maxbibnames=10,url=false,doi=true,backref=false]{biblatex}

\renewbibmacro{in:}{\ifentrytype{article}{}{\printtext{\bibstring{in}\intitlepunct}}}

\makeatletter
\DeclareFontFamily{U}{tipa}{}
\DeclareFontShape{U}{tipa}{m}{n}{<->tipa10}{}
\newcommand{\arc@char}{{\usefont{U}{tipa}{m}{n}\symbol{62}}}%

\newcommand{\arc}[1]{\mathpalette\arc@arc{#1}}

\newcommand{\arc@arc}[2]{%
  \sbox0{$\m@th#1#2$}%
  \vbox{
    \hbox{\resizebox{\wd0}{\height}{\arc@char}}
    \nointerlineskip
    \box0
  }%
}
\makeatother

\usepackage{pgf}

\numberwithin{equation}{subsection}

\begin{document}
\title{Mathematical model of Nucleocytoplasmic Transport and Nuclear-to-Cell Ratio in a growing cell}

\author{Xuesong Bai, Thomas G. Fai}
\address{Department of Mathematics, Brandeis University, Waltham, MA}

\date{\today}

\begin{abstract} 
It has been observed that the growth of the nucleus and the cytoplasm is coordinated during cell growth, resulting in a nearly constant nuclear-to-cell volume ratio (N/C) throughout the cell cycle. Previous studies have shown that the N/C ratio is determined by the ratio between the number of proteins in the nucleus and the total number of proteins in the cell. These observations suggest the importance of the nucleocytoplasmic transport process in nuclear size by regulating protein concentrations in the nucleus and cytoplasm. This paper combines a biophysical model of Ran-mediated nucleocytoplasmic transport and a simple cell growth model to provide insights into several key aspects of the N/C ratio homeostasis in growing cells. Our model shows that the permeability of the nuclear envelope needs to grow in line with the cell to maintain a nearly constant N/C ratio, that several parameters involved in the nucleocytoplasmic transport mechanism and gene translation significantly affect the N/C ratio, and that Ran may potentially compensate for the lack of NTF2 in the nucleocytoplasmic transport mechanism to maintain a viable N/C ratio. However, this compensation is possible only if RanGDP is allowed to translocate through the nuclear envelope independently of NTF2.
\end{abstract}

\maketitle

\section{Introduction}

It has been observed that the growth of the nucleus and the cytoplasm are coordinated during cell growth, resulting in a nearly constant nuclear-to-cell volume ratio (N/C ratio) in specific cell types such as fission yeast \cite{conklin1914cell,moore2019determination,neumann2007nuclear}.
Recent work has found that, in fission yeast, this volume ratio is primarily determined by the osmotic balance across the nuclear envelope, which in turn is determined by the ratio between the number of proteins in the nucleus and the total number of proteins in the cell \cite{deviri2022balance,lemiere2022control}.

In our previous work \cite{bai2025stochastic}, we studied the effects of fluctuations on the homeostasis of the protein number ratio. We used a simplified stochastic gene translation model in which the translation rate of proteins is proportional to their relative gene fractions to study the effects of intrinsic noise arising from the translation process and extrinsic noise arising from the random partitioning of molecules at cell division. We showed that homeostasis of the protein number ratio is robust to these fluctuations and that the fluctuations become negligible as the protein numbers become sufficiently large. However, our previous model portrays a simplified scenario in which the nuclear proteins are defined as proteins with nuclear localization sequences (NLSs) without regard to their actual location in the cell, and we have not considered how they are imported into the nucleus; that is, the process of nucleocytoplasmic transport.

Coordination between several biological processes is required for large molecules, such as proteins, to enter the nucleus. The Ran-mediated nucleocytoplasmic transport process has been well characterized experimentally and described in terms of existing theoretical models (e.g., \cite{macara2001transport,weis2002nucleocytoplasmic,zilman2007efficiency}). Several mathematical models of this process, with different modeling assumptions and various levels of complexity, have been developed in previous research. For example, in \cite{timney2006simple}, the authors formulated a simple pump-leak model involving only cargo proteins and importins and fit to \textit{in vivo} import rate data. In \cite{kim2013simple}, the authors studied a more biologically detailed model taking into account the dynamics of Ran protein, and in \cite{wang2017thermodynamic}, the authors added further details into the model, such as the role of NTF2 in the Ran cycle. More detailed models exploring other aspects of the nucleocytoplasmic transport process, such as the roles of different types of importins (Imp\textalpha, Imp\textbeta) \cite{riddick2005systems} or of detailed reactions in the RanGDP-RanGTP phosphorylation \cite{gorlich2003characterization} are also studied. While these mathematical models provide various insights into the nucleocytoplasmic transport process, to our knowledge, none of these existing models has considered the effect of cell growth in that they all assume cells are static instead of growing.

That is, while cell growth models that incorporate the nucleocytoplasmic transport process do exist \cite{leech2022mathematical,wu2022correlation}, these models greatly simplify the details of the transport process. While the simplification provides benefits such as a reduced number of parameters, these simplified models do not aim to capture the detailed dynamics in the Ran-mediated nucleocytoplasmic transport process.

This paper aims to build upon these previous models to provide a mechanistic model of nucleocytoplasmic transport and cell growth and identify the key aspects of the nucleocytoplasmic transport process that affect the N/C ratio. We combine our previous simple cell growth model \cite{bai2025stochastic} with a detailed biophysical model of nucleocytoplasmic transport \cite{wang2017thermodynamic} to derive a model describing the dynamics of nuclear cargo proteins, importins, Ran proteins, NTF2 proteins, together with the dynamics of ribosomes and cytoplasmic proteins in growing cells.

We face several challenges in modeling and studying the nucleocytoplasmic transport process in growing cells. First, the cell volume and the nuclear envelope's permeability may change over time rather than staying constant. We use a simple linear model for the dependence of the nuclear envelope permeability on cell volume, which may be viewed as a linear approximation to a more complicated dependence on volume. Even within this simple linear regime, there are several possible models for the permeability of the nuclear envelope \cite{leech2022mathematical}. We must closely examine these modes and study their implications on the N/C ratio. Second, the nucleocytoplasmic transport model contains a relatively large number of parameters, making it difficult to constrain the parameters and yield easily interpretable results. Finally, as explained later, we must modify the original model in \cite{wang2017thermodynamic} to explain specific experimental observations.

The main insights of this study are as follows: (i) We show that in order for the system to have a stable steady state so that the N/C ratio approaches a constant value, the nuclear envelope permeability must scale with the number of cytoplasmic \textit{housekeeping} proteins or with the volume of the nucleus; (ii) we perform a sensitivity analysis of the N/C ratio, focusing on the parameters governing the nucleocytoplasmic transport process, and provide an interpretation of these results; (iii) we show that increasing the expression of Ran may relieve the requirement on NTF2 for maintaining a normal N/C ratio homeostasis and that this compensation effect of Ran requires the NTF2-independent permeability of RanGDP across the nuclear envelope.

\section{Model of Ran-mediated nucleocytoplasmic transport and cell growth}
Our model is based on a previous biophysical model of Ran-mediated nucleocytoplasmic transport \cite{wang2017thermodynamic}, which is combined with a simple cell growth model that we proposed in our previous work \cite{bai2025stochastic}.

The nucleocytoplasmic transport model, detailed in \cite{wang2017thermodynamic}, consists of 11 basic reactions, and depicts the transport of cargo proteins through binding with importins, the unloading of cargo proteins facilitated by the binding and unbinding of RanGTP with importins, and the consumption of energy through the conversion between RanGTP and RanGDP. The model also incorporates the diffusion of the proteins and protein complexes in the nucleocytoplasmic transport machinery across the nuclear envelope.

The cell growth model in \cite{bai2025stochastic} depicts the exponential growth of proteins via autocatalytic replication of ribosomes.

A summary of modifications applied to the original model in \cite{wang2017thermodynamic} is given in Appendix \ref{modelcomparison}.

\begin{figure}[h]
 \includegraphics[width=0.8\textwidth]{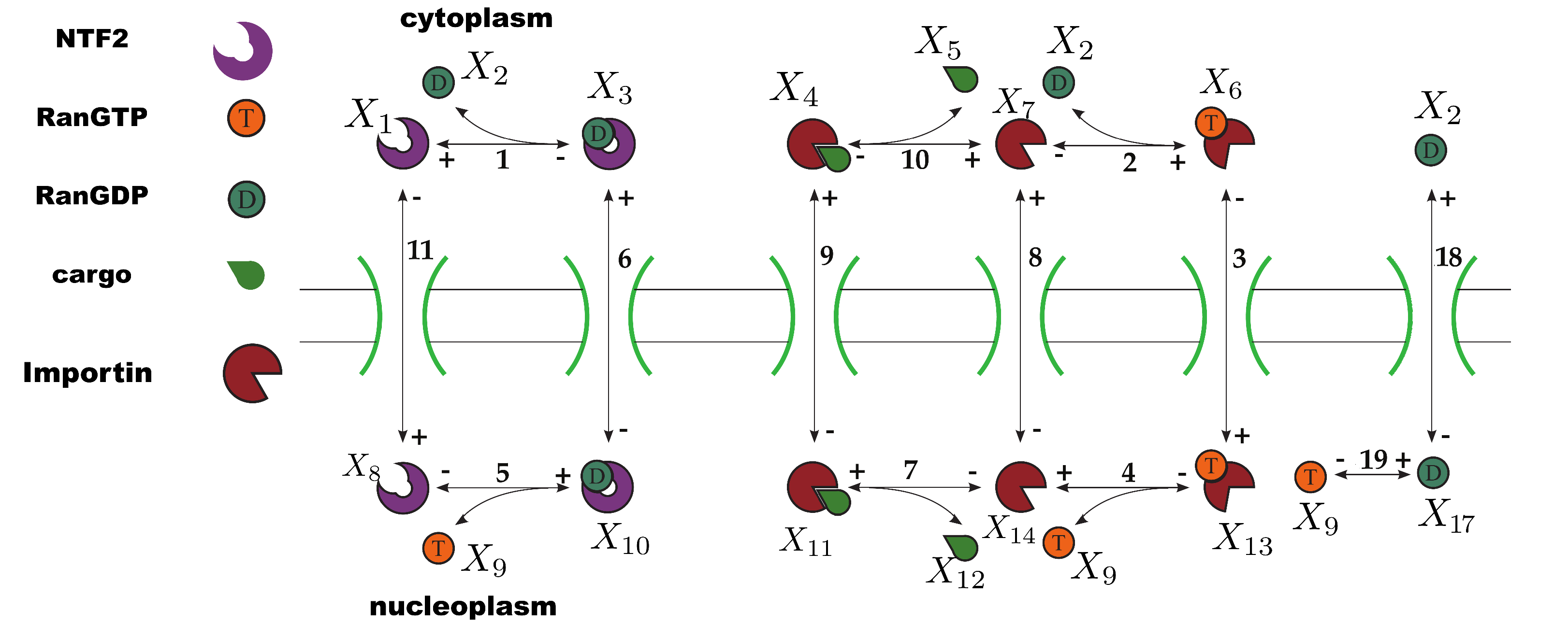}
 \caption{Species and reactions involved in the nucleocytoplasmic transport machinery. Arrows correspond to reactions, with their respective labels corresponding to the equation numbers provided in Appendix \ref{reactionlist}. Adapted from Fig. S1 in \cite{wang2017thermodynamic} and available under CC BY license
.}
 \label{Machinery}
\end{figure}

\subsection{Components of the model}
Our model has three major parts. The first part of the model consists of reactions (\ref{R1}) to (\ref{R11}), which are adapted from the model detailed in \cite{wang2017thermodynamic} and gives a dynamical description of the nucleocytoplasmic transport process. These reactions can be further categorized into three types \cite{biswas2023conserved}: binding and unbinding in the cytoplasm ((\ref{R1}), (\ref{R2}), (\ref{R10})); binding and unbinding in the nucleus ((\ref{R4}), (\ref{R5}), (\ref{R7})); diffusion through the nuclear pore complexes (NPCs) ((\ref{R3}), (\ref{R6}), (\ref{R8}), (\ref{R9}), (\ref{R11})).

The second part of the model consists of reactions (\ref{R12}) to (\ref{R17}), which are adapted from our previous cell growth model \cite{bai2025stochastic} but are modified to include more biomolecule species and describes the autocatalytic biogenesis of the ribosomes and the synthesis of other proteins.

The third part of the model consists of reactions (\ref{R18}) and (\ref{R19}), which are additional reactions in the nucleocytoplasmic transport process \cite{riddick2005systems} that will be discussed in Section \ref{compensation}.

A schematic of all the species and reactions in the nucleocytoplasmic transport machinery, which corresponds to the first and third part of our model, is shown in Fig. \ref{Machinery}.

\subsection{Modeling assumptions}\label{assumptions}
For the first part of the model, we proceed as in \cite{wang2017thermodynamic}, with some crucial adjustments to account for the changes in a growing cell, as we describe next. For the second part of the model, our basic modeling assumptions are that the protein numbers can grow without any restriction, and that the translation rate of protein species $j$ is proportional to its gene fraction $\phi_j$ \cite{bai2025stochastic}.

\subsubsection{Dynamics of the RanGAP- and RanGEF-mediated reactions}
In the original version of the nucleocytoplasmic transport model, the rates of the RanGAP-mediated reaction (\ref{R2}) and the RanGEF-mediated reaction (\ref{R5}) are assumed to be constant. In our model of growing cells, however, the numbers of RanGAP and RanGEF may also be increasing; as a result, we must check whether the reaction rates mentioned above are still constant. To answer this question, we take a closer look at the two reactions (\ref{R2}) and (\ref{R5}).

According to (S38) and (S39) in \cite{wang2017thermodynamic}, the RanGAP-mediated reaction (\ref{R2}) consists of the following two reactions:
\begin{subequations}
\begin{align}
\label{GAP1} IRan_{Tc}+GDP &\xrightleftharpoons[k_\gamma^-]{k_\gamma^+} IRan_{Dc}+GTP; \\
\label{GAP2} IRan_{Dc} &\xrightleftharpoons[k_\delta^-]{k_\delta^+} I_c+Ran_{Dc}.
\end{align}
\end{subequations}
Reaction (\ref{GAP1}) is mediated by RanGAP so that the reaction rate coefficients $k_\gamma^+$ and $k_\gamma^-$ depend on the number of RanGAP proteins. If we assume that both reaction rates are proportional to the number of RanGAP, then the ratio $k_\gamma^+/k_\gamma^-$ will remain constant despite any increase in the number of RanGAP. Reaction (\ref{GAP2}) does not depend on RanGAP, so the reaction rate coefficients $k_\delta^+$ and $k_\delta^-$ are constant. Therefore, if the concentrations $[GDP]$ and $[GTP]$ of free GDP and GTP, resp., remain constant, then at a steady state, the ratio
\begin{equation}
 \frac{k_2^+}{k_2^-} = \frac{k_\gamma^+}{k_\gamma^-}\frac{k_\delta^+}{k_\delta^-}\frac{[GDP]}{[GTP]}
\end{equation}
also remains constant during growth, and the propensity functions for reaction (\ref{R2}) in our growth model remain the same as those in the original model.

Similarly, according to (S30) and (S31), the RanGEF-mediated reaction (\ref{R5}) is also a two-step reaction:
\begin{subequations}
\begin{align}
\label{GEF1} NRan_{Dn}+GTP &\xrightleftharpoons[k_\alpha^-]{k_\alpha^+} NRan_{Tn}+GDP; \\
\label{GEF2} NRan_{Tn} &\xrightleftharpoons[k_\beta^-]{k_\beta^+} N_n+Ran_{Tn}.
\end{align}
\end{subequations}
RanGEF only mediates reaction (\ref{GEF1}), and by an analogous assumption that $k_\alpha^+$ and $k_\alpha^-$ are both proportional to the number of RanGEF, the ratio $k_\alpha^+/k_\alpha^-$ remains constant despite the growth of the number of RanGEF; therefore, the ratio
\begin{equation}
 \frac{k_5^+}{k_5^-} = \frac{k_\alpha^+}{k_\alpha^-}\frac{k_\beta^+}{k_\beta^-}\frac{[GTP]}{[GDP]}
\end{equation}
remains constant during growth, and the propensity functions for reaction (\ref{R5}) in our growth model remain the same as those in the original model.

We note that in this paper, we do not attempt to model the dynamics of (\ref{R2}) and (\ref{R5}) in microscopic detail; instead, we assume coarse-grained and phenomenological dynamics. One could, in principle, consider more detailed models of the RanGAP- and RanGEF-mediated reactions, which is beyond the scope of the current paper.

\subsubsection{Determination of nuclear and cytoplasmic volume}
Consider the total number of proteins in the cytoplasm and the nucleus, respectively:
\begin{equation*}
 P_{cyto} = N_c+Ran_{Dc}+NRan_{Dc}+IP_{nc}+P_{nc}+IRan_{Tc}+I_c+R_c+P_c,
\end{equation*}
\begin{equation*}
 P_n = N_n+Ran_{Tn}+Ran_{Dn}+NRan_{Dn}+IP_{nn}+P_{nn}+IRan_{Tn}+I_n.
\end{equation*}
Similar to the assumption on the cytoplasmic volume  invoked in \cite{wu2022correlation}, we assume that the cytoplasmic volume is proportional to the total number of cytoplasmic proteins, plus a minimum  (non-osmotic) volume:
\begin{equation}\label{Vc}
 V_{cyto} = C_0+C_1P_{cyto}.
\end{equation}
Eq. (\ref{Vc}) may be viewed as a first-order approximation of a more complicated function that describes the dependency of the cytoplasmic volume on the cytoplasmic protein number. In our current model, $C_0$ is set to be a fixed minimum volume, and $C_1$ is determined by the initial cytoplasmic volume $V_{cyto,0}$ and the initial number of cytoplasmic proteins $P_{cyto,0}$.

For the nuclear volume, we follow previous studies on the N/C ratio \cite{deviri2022balance,lemiere2022control} and assume that the osmotic balance across the nuclear envelope determines the nuclear volume:
\begin{equation}\label{Vn}
\frac{V_n}{V_{cyto}} = \frac{P_n}{P_{cyto}}:=\Phi_{NCyto}\quad\mbox{so that}\quad V_n = \frac{P_n}{P_{cyto}}V_{cyto}.
\end{equation}
Here $\Phi_{NCyto}$ is the nuclear-to-cytoplasmic ratio. For the rest of the paper, the N/C ratio is defined to be the nuclear-to-cell ratio $\Phi_{NCell}$. We derive from (\ref{Vn}) that
\begin{equation}\label{NCRatio}
\Phi_{NCell} := \frac{V_n}{V_n+V_{cyto}} = \frac{P_n}{P_n+P_{cyto}}.
\end{equation}

\subsubsection{Permeability of the nuclear envelope}\label{permeability}
In the nucleocytoplasmic transport process, there are several steps (reactions (\ref{R3}), (\ref{R6}), (\ref{R8}), (\ref{R9}), and (\ref{R11})) that involve proteins or protein complexes to translocate through the NPCs, and the rate at which these molecules pass through the NPCs significantly affects the rate of nucleocytoplasmic transport. We will not model the microscopic dynamics of the NPCs (see, e.g., \cite{moussavi2011brownian} for a detailed model of the NPCs); instead, we model the overall permeability of the nuclear envelope. 
\begin{figure}[h]
 \includegraphics[width=0.7\textwidth]{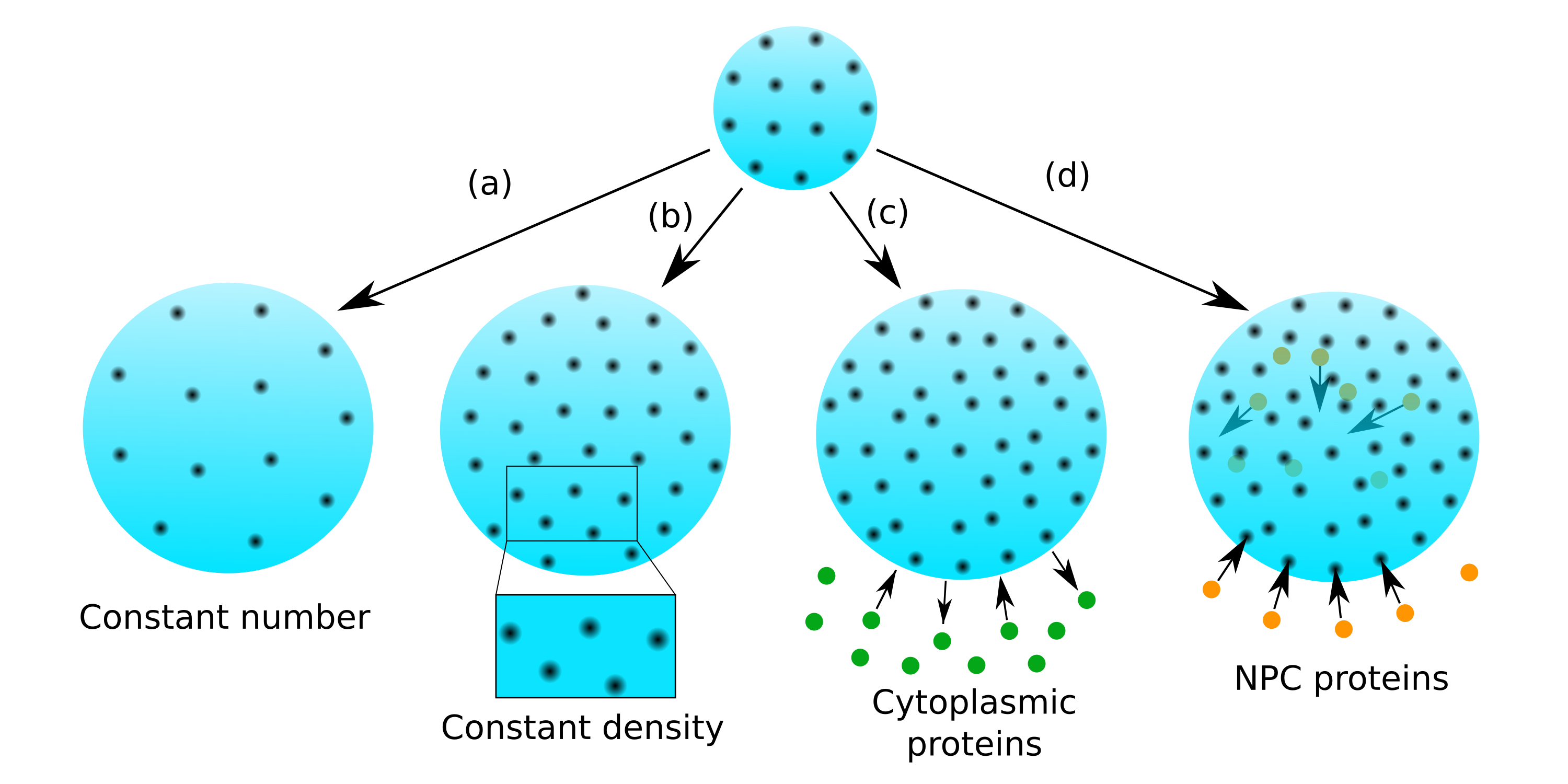}
 \caption{Different assumptions on the nuclear envelope permeability during cell growth. (a) constant number of NPCs; (b) the number of NPC scales with surface area; (c) the number of NPC scales with the number of cytoplasmic proteins; (d) the number of NPC scales with the number of nuclear proteins, which is equivalent to scaling with the nuclear volume in our model.}
 \label{NEPermeability1}
\end{figure}

In the original version of the nucleocytoplasmic transport model \cite{wang2017thermodynamic}, the permeability of the nuclear envelope $a$ is assumed to be constant (Fig. \ref{NEPermeability1} (a)). That is,
\begin{equation}\label{constperm}
 a = a_0.
\end{equation}
In our model of growing cells, the volume of the nucleus is growing, and as a result, the surface area of the nuclear envelope is increasing. In addition, new proteins that form the NPCs may be synthesized to increase the number of NPCs. Therefore, the overall permeability $a$ of the nuclear envelope should also increase with the cell growth. We assume that the permeability of the nuclear envelope is proportional to the total number of NPCs. Several possible models exist of how the number of NPCs changes over time. First, experimental evidence shows that the NPCs have a roughly constant density on the nuclear membrane (Fig. 1C in \cite{dultz2010live}). In this case, we may assume that the permeability is proportional to the surface area of the nuclear envelope:
\begin{equation}\label{surfaceareascaling}
 a = C_{np}V_n^\frac23.
\end{equation}
As a second possibility, if the NPCs on the nuclear membrane are rapidly exchanged with a cytoplasmic pool of NPC proteins so that the total number of NPCs is proportional to the number of cytoplasmic \textit{housekeeping} proteins $P_c$, we may assume that the permeability is also proportional to $P_c$ \cite{wu2022correlation}:
\begin{equation}\label{Pcscaling}
 a = C_{np}P_c.
\end{equation}
Finally, if the NPC proteins are first imported into the nucleus and then inserted into the nuclear membrane so that the total number of NPCs is proportional to the total number of nuclear proteins $P_n$, we may assume that the permeability is also proportional to $P_n$. Equivalently, by our assumption (\ref{Vn}) that the nuclear volume $V_n$ scales with $P_n$,\footnote{Strictly speaking, the scaling factor $V_{cyto}/P_{cyto}$ is not constant because of (\ref{Vc}), but this is a good approximation in practice.} we may assume that the permeability is proportional to the nuclear volume $V_n$:
\begin{equation}\label{Vnscaling}
 a = C_{np}V_n.
\end{equation}
\begin{figure}[h]
 \includegraphics[width=0.5\textwidth]{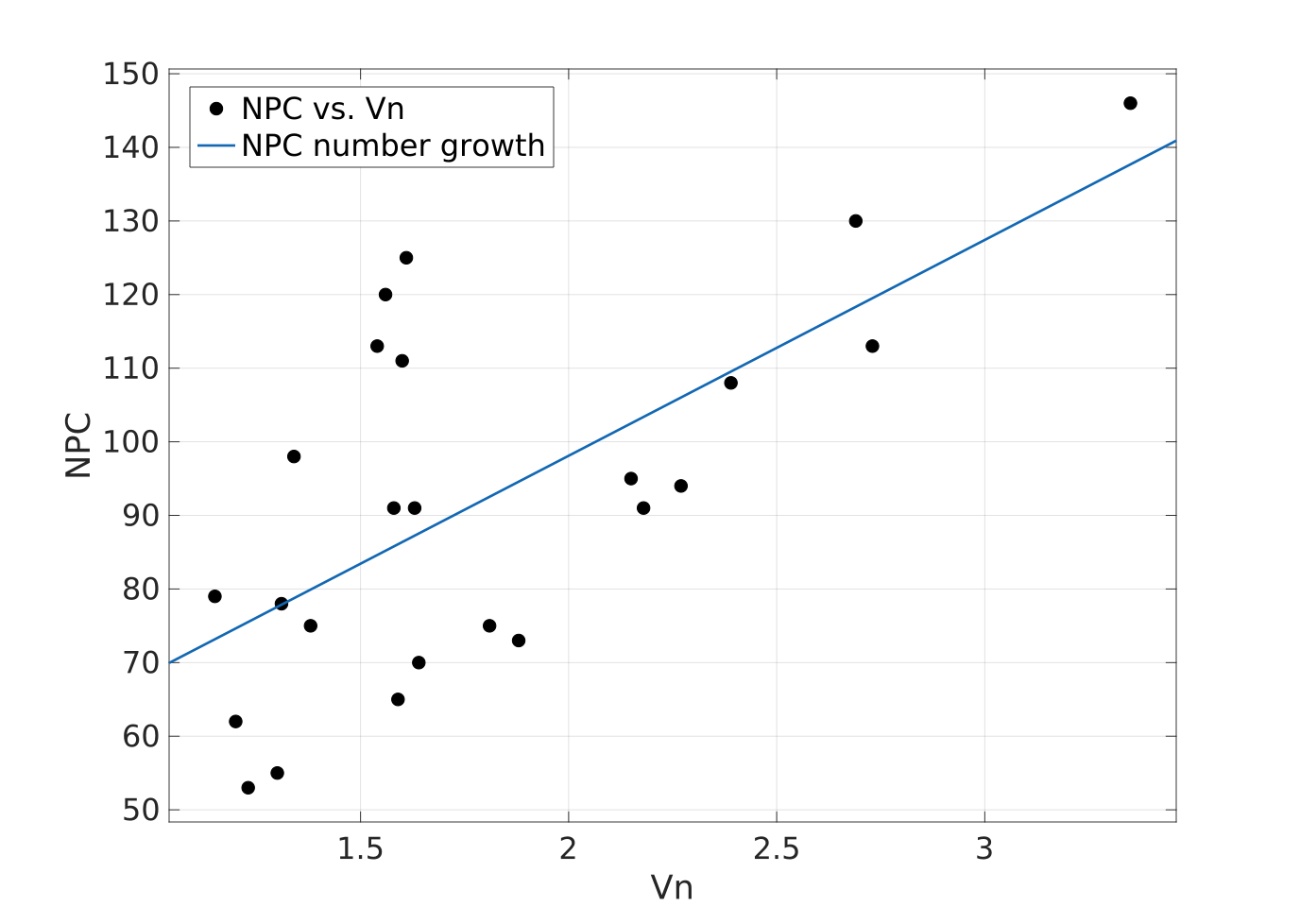}
 \caption{Linear fit using data of NPC number and nuclear volume (measured in $\mu$m$^3$) from \cite{winey1997nuclear}.}
 \label{NPC_vs_Vn}
\end{figure}
There is also experimental evidence in support of (\ref{Vnscaling}). In \cite{winey1997nuclear}, the authors reconstructed 3D models of yeast (\textit{Saccharomyces cerevisiae}) nuclei from electron micrographs and measured the NPC number and nuclear dimensions for each nucleus. Fig. \ref{NPC_vs_Vn} shows the linear fit to the data on NPC number and nuclear volume provided by Table 1 in \cite{winey1997nuclear}. While the data points are scattered, we see a clear positive correlation between nuclear volume and NPC number.

Given the complexity of the NPC assembly process and its regulation \cite{dultz2008systematic,guglielmi2020nuclear,otsuka2018mechanisms}, the dynamics of NPC number and nuclear envelope permeability may vary among different cell types. As we will see later, the three possible models of nuclear envelope permeability presented above have various implications on the N/C ratio homeostasis, which allows us to exclude some of these possibilities by showing that they are incompatible with experimental observations.

\subsection{Propensity functions and the ODE system}
The complete system of ordinary differential equations (ODEs) describing the dynamics of the model may be expressed as
\begin{equation}\label{ODEsystem}
 \frac{d\mathbf{x}}{dt} = \mathbf{M}{\bm\lambda}(\mathbf{x}).
\end{equation}
Here $\mathbf{M}$ is the stoichiometric matrix corresponding to the reactions listed in Appendix \ref{reactionlist}, $\mathbf{x}$ is the variable vector whose components are listed in Appendix \ref{vlist}, and ${\bm\lambda}(\mathbf{x})$ is the propensity function vector whose components are listed in Appendix \ref{plist}. Note that each component of $\mathbf{x}$ keeps track of the \textit{number} of the corresponding protein species. We can also calculate the \textit{concentration} $\mathbf{c}$ of the protein species from $\mathbf{x}$:
\begin{equation}\label{concentration}
 c_i = \frac{x_i}{N_AV_{cyto}}\quad\mbox{(Cytoplasmic species)}\quad\mbox{or}\quad c_i = \frac{x_i}{N_AV_n}\quad\mbox{(Nuclear species)},
\end{equation}
where $N_A=6.02214076\times 10^{23}\mathrm{mol}^{-1}$ is the Avogadro constant, and $V_{cyto}$ and $V_n$ are given by (\ref{Vc}) and (\ref{Vn}), respectively.

We solve (\ref{ODEsystem}) numerically to obtain time evolutions $\mathbf{x}(t)$. The ODE system (\ref{ODEsystem}) is stiff. Therefore, we use solvers for stiff equations such as Matlab's \texttt{ode15s} or \texttt{ode23s}.

\subsection{Number of proteins vs concentration of proteins, and steady states of the system}
In the original nucleocytoplasmic transport model \cite{wang2017thermodynamic} of cells without growth, keeping track of the \textit{number} of proteins is effectively equivalent to keeping track of the \textit{concentration} of proteins because there is no protein synthesis, and the cell volume is set to be constant. The number and concentration of each protein species will eventually reach a steady state.

By contrast, in our model of growing cells, the number of proteins is increasing due to the autocatalytic growth of ribosomes (\ref{R12}). As a result, the cell volume is growing. Since we assume the growth is unrestricted in Section \ref{assumptions}, each of the protein numbers $x_i$ will grow towards infinity. Therefore, the ODE system (\ref{ODEsystem}) that keeps track of the \textit{number} of proteins does not have a steady state. That is, $\mathbf{x}$ \textit{does not approach} any finite $\mathbf{x}^*$ as $t\to\infty$. However, if we switch our viewpoint from the \textit{number} of proteins $x_i$ to the \textit{concentration} of proteins $c_i$, where $c_i$ is defined by (\ref{concentration}), we see that the concentrations $c_i$ will reach steady states even if the cell is growing; that is, $\mathbf{c}\to\mathbf{c}^*$ as $t\to\infty$.  For the rest of this paper, we define the steady state of the system as the condition that the concentration of proteins $\mathbf{c}$ reaches the steady state $\mathbf{c}^*$.

\subsection{Parameter selection}
For the nucleocytoplasmic transport part of the model, the numerical values of all reaction rate parameters in reaction (\ref{R1}) - (\ref{R11}) are taken from the original model \cite{wang2017thermodynamic}. For the cell growth part of the model, the translation rate $k_t$ is set to be $0.005$, and the gene fraction of ribosomes is set to be $\phi_r = 0.02$; both are taken from our previous cell growth model \cite{bai2025stochastic}. Other gene fractions are calculated from simulations of the original nucleocytoplasmic transport model without cell growth, in which the N/C ratio is roughly $0.08$. For each protein species, the gene fraction is set to be the fraction of the protein in the proteome at the steady state. We set the initial cytoplasmic volume for all the simulations to $V_{cyto,0} = 500\mathrm{\mu m}^3$.

\section{Results}


Matlab codes for the following results are available on GitHub\footnote{\href{https://github.com/topgunbai683/nucleocytoplasmic-transport-model.git}{https://github.com/topgunbai683/nucleocytoplasmic-transport-model.git}}.

\subsection{Long-term behaviors of the N/C ratio and nuclear envelope permeability}
We first examine how the different models of the nuclear envelope permeability listed in Section \ref{permeability} lead to different long-term behaviors of the N/C ratio. In particular, we intend to discover which permeability model lead to a constant and nonzero N/C ratio. Stated in mathematical terms, the question is to determine permeability functions $a=a(\mathbf{x})$ so that the N/C ratio $\Phi_{NCell}(\mathbf{x})\to\Phi_{NCell}^*>0$ as $t\to\infty$.

We set the initial conditions so that the value for the initial nuclear-to-cytoplasmic volume ratio is $\Phi_{NCyto,0} = 0.0886$, and the initial nuclear volume is then $V_{n,0} = \Phi_{NCyto,0}V_{cyto,0}$. For the constant permeability case (\ref{constperm}), we set the value of the nuclear permeability to be $a_0 = 100\mathrm{\mu m}^3/\mathrm{s}$. For the two permeability functions (\ref{surfaceareascaling}) and (\ref{Pcscaling}), we set the constant $C_{np}$ to be such that the nuclear envelope permeability $a$ is equal to $a_0$ when $V_n = V_{n,0}$ and $P_c = P_{c,0}$, respectively.
\begin{figure}[h]
 \includegraphics[width=0.4\textwidth]{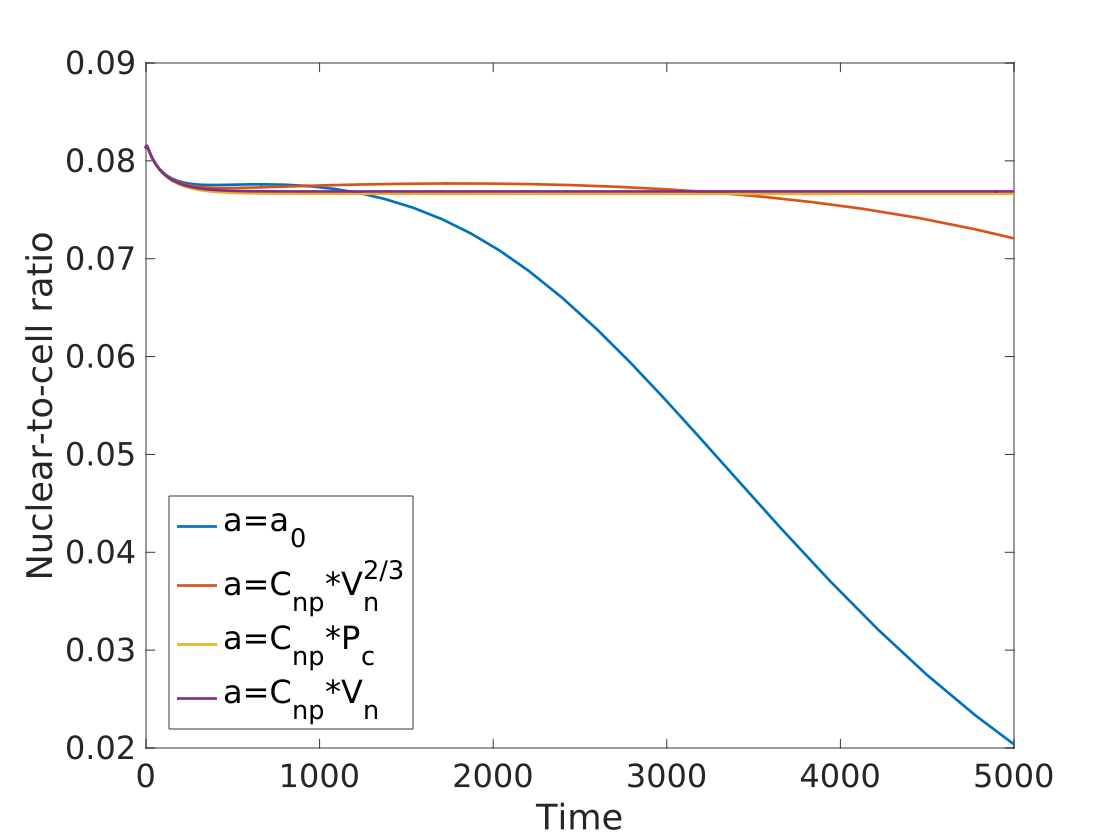}
 \caption{Long-term behaviors of the nuclear-to-cell ratio under different models of the nuclear envelope permeability.}
 \label{NC_steady_state}
\end{figure}

Fig. \ref{NC_steady_state} shows the simulation results of the N/C ratio dynamics obtained using different nuclear envelope permeability models. The constant nuclear envelope permeability model (\ref{constperm}) cannot keep up with the cell growth, and the N/C ratio decreases relatively quickly. This result is inconsistent with experimental observations \cite{lemiere2022control}. Therefore, the constant permeability model (\ref{constperm}) is ruled out. The decrease in the N/C ratio is not unexpected if we consider the propensity functions for diffusion through NPCs (for example, (\ref{R3})) in Table \ref{propensitylist}. If $a(\mathbf{x})=a_0$ and the volumes $V_{cyto}$ and $V_n$ keep increasing, then
\begin{equation*}
 \frac{a(\mathbf{x})}{V_n(\mathbf{x})}=\frac{a_0}{V_n(\mathbf{x})}\to 0\quad\mbox{and}\quad\frac{a(\mathbf{x})}{V_{cyto}(\mathbf{x})}=\frac{a_0}{V_{cyto}(\mathbf{x})}\to 0\quad\mbox{as}\quad V_n(\mathbf{x}), V_{cyto}(\mathbf{x}) \to \infty.
\end{equation*}
So, the diffusion rate through NPCs eventually decreases to 0, resulting in the accumulation of nuclear cargo proteins in the cytoplasm.

For the permeability model (\ref{surfaceareascaling}) that scales with the surface area of the nuclear envelope, the N/C ratio remains close to the expected value, about 8\%, for a reasonably long time but eventually begins to decrease. The decrease in the N/C ratio may be explained again by the propensity functions for diffusion through NPCs. If $V_n$ keeps increasing, then
\begin{equation*}
 \frac{a(\mathbf{x})}{V_n(\mathbf{x})}=\frac{C_{np}V_n(\mathbf{x})^\frac23}{V_n(\mathbf{x})}=\frac{C_{np}}{V_n(\mathbf{x})^\frac13}\to 0\quad\mbox{as}\quad V_n(\mathbf{x}) \to \infty,
\end{equation*}
so the diffusion rate through NPCs eventually decreases to 0, as in the previous case.

Only the permeability models (\ref{Pcscaling}) that scale with the number of cytoplasmic housekeeping proteins and (\ref{Vnscaling}) that scales with the nuclear volume will eventually lead to steady states of the N/C ratio homeostasis in the presence of growth. We see from Fig. \ref{NC_steady_state} that the steady states of the N/C ratio under (\ref{Pcscaling}) and (\ref{Vnscaling}) are very close to each other and nearly indistinguishable. Given the above observations, in the following simulations, we use the nuclear envelope permeability function (\ref{Vnscaling}) based on the data in \cite{winey1997nuclear} shown in Fig. \ref{NPC_vs_Vn}. 

Note that the volumes $V_n(\mathbf{x}), V_{cyto}(\mathbf{x}) \to \infty$ is the consequence of the unrestricted growth assumption specified in Section \ref{assumptions}. Although here for simplicity we have chosen not to model the division process, in reality cells of course divide and do not grow without bound. Therefore, the volumes described by the model may be considered as belonging to a population rather than an individual cell. 

Also due to the fact that cells do not grow without restrictions and may eventually divide, it is possible that the nuclear envelope permeability function (\ref{surfaceareascaling}) is sufficient for maintaining a near constant N/C ratio during one cell cycle and, therefore, cannot be ruled out definitively. Further studies of the nuclear envelope permeability in growing cells are needed to understand the mode of NPC scaling in cells.

\subsection{Stable steady states and N/C ratio homeostasis}
We next test whether the steady state of the system is stable so that the N/C ratio approaches the constant value homeostatically. That is, we test whether the N/C ratio will reach the same constant value starting from different initial conditions.
\begin{figure}[h]
 \includegraphics[width=0.4\textwidth]{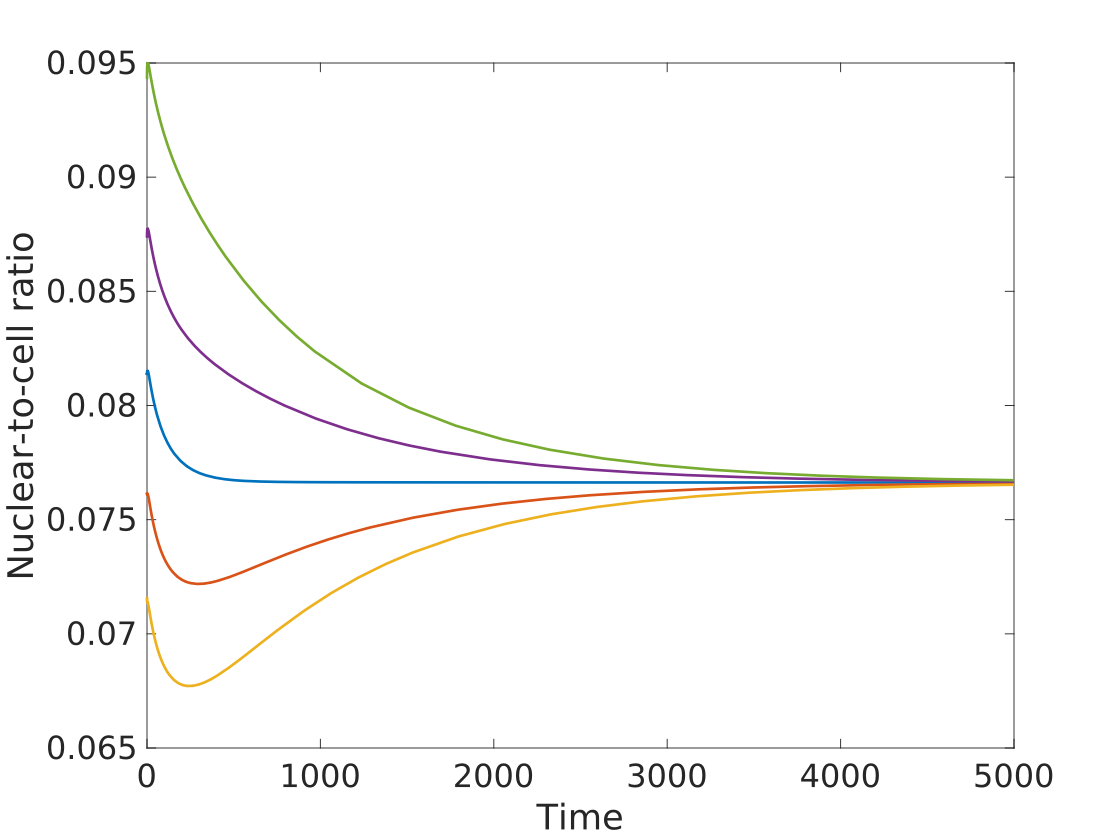}
 \caption{Long-term behaviors of the nuclear-to-cell ratio, starting from different initial conditions.}
 \label{NC_homeostasis}
\end{figure}

Fig. \ref{NC_homeostasis} shows that different initial conditions of the N/C ratio eventually lead to the same constant value $\Phi_{NCell}^*$. This phenomenon means that the steady state $\mathbf{c^*}$ of the system is stable and that perturbations in the N/C ratio are corrected as the cell grows. Consequently, an N/C ratio homeostasis exists. Note that there may be an initial transient during which the N/C ratio temporarily goes further away from the steady state before eventually returning to the steady-state value.

Moreover, we see from Fig. \ref{NC_homeostasis} that, generally speaking, the further away the initial condition is from $\Phi_{NCell}^*$, the faster the N/C ratio curve back toward $\Phi_{NCell}^*$. This observation is consistent with our previous models of N/C ratio homeostasis during exponential growth \cite{bai2025stochastic,lemiere2022control}.

\subsection{Sensitivity analysis}\label{sensitivity}
Next, we perform a sensitivity analysis to determine which parameters in the model are crucial to determining the N/C ratio. The quantity of interest (QoI) is set to be the N/C ratio $\Phi_{NCell}$ at the steady state $\mathbf{c^*}$ of the system.

For the sensitivity analysis, we focus on the parameters governing the nucleocytoplasmic transport process. We will assume that the two cytoplasmic volume parameters $C_0$ and $C_1$, the GTP/GDP ratio $\theta$, as well as the gene fractions of the ribosomes and nuclear proteins, $\phi_r$ and $\phi_n$, are constant. At the same time, we perform sensitivity analysis on all the other parameters. In reality, changes in the abovementioned variables may affect the N/C ratio. However, when choosing the parameters of interest, we will restrict ourselves to the parameters directly related to the nucleocytoplasmic transport process, except for the translation rate $k_t$.

We use Spearman's correlation coefficient and partial rank correlation coefficient (PRCC) for the sensitivity analysis \cite{cogan2022mathematical,marino2008methodology}, and we use Matlab's \texttt{lhsdesign} to generate Latin hypercube sample matrices of the parameters. The hypercube has sides that are $\pm10\%$ of the nominal values of the parameters to avoid pushing the parameters far outside the initial parameter regime.
\begin{figure}[h]
 \includegraphics[width=\textwidth]{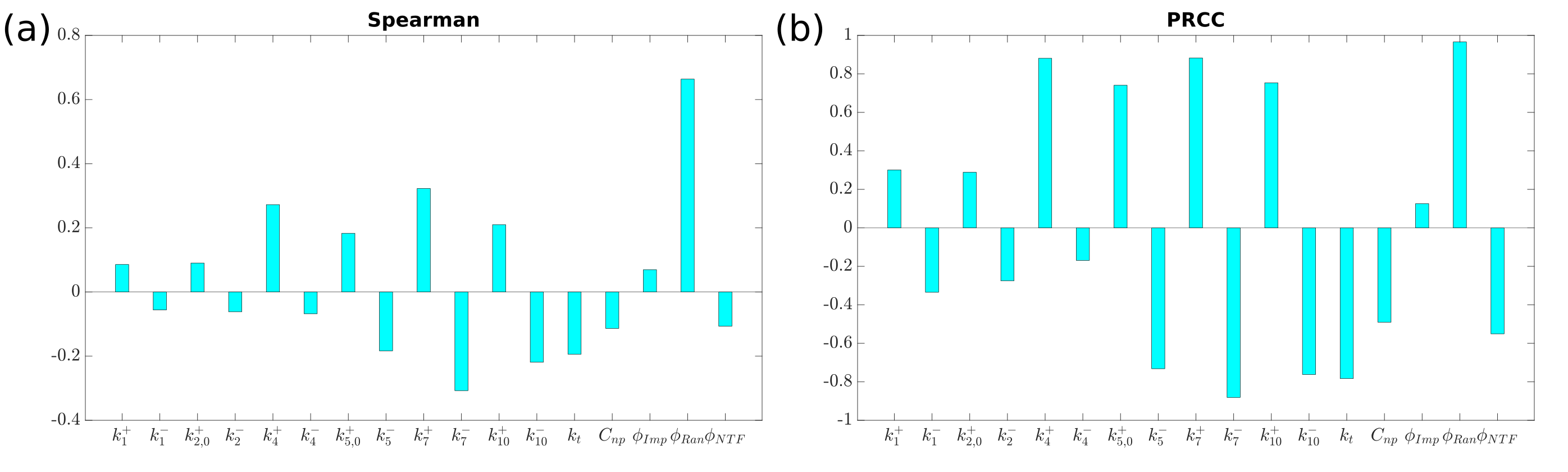}
 \caption{Sensitivity analysis of the parameters of interest. QoI is the N/C ratio $\Phi_{NCell}$ (a) Spearman correlation; (b) Partial rank correlation.}
 \label{Sensitivity}
\end{figure}

Fig. \ref{Sensitivity} shows the Spearman and partial rank correlation coefficients and partial rank correlation coefficients between the N/C ratio and each parameter of interest. Intuitively, if a particular correlation coefficient is positive (negative), increasing the parameter will lead to an increase (decrease) in the N/C ratio, and the magnitude of the correlation coefficient shows how strongly the N/C ratio is correlated to the parameter. We see from Fig. \ref{Sensitivity} that, while the sensitivity analysis results from the two methods are not identical, if we compare the effects of the various parameters on the N/C ratio, both methods show the same pattern. Therefore, the sensitivity analysis result is robust with respect to the choices of methods. Having established this robustness to the particular method used, for convenience we will use the PRCC methods in the following sensitivity analyses.

We then take a deeper look into the results of the above sensitivity analysis by performing a further sensitivity analysis to study how variations of selected parameters affect the localization of proteins in the nucleocytoplasmic transport machinery. In particular, we choose the quantities of interest to be the following fractions of protein numbers in the nucleus at the steady state $\mathbf{c^*}$ of the system:
\begin{enumerate}
 \item Fraction of all nucleocytoplasmic transport machinery proteins in the nucleus:
 \begin{equation*}
 \Phi_{NCTn} = \frac{NCT_{n}}{NCT_{n}+NCT_{c}}
 \end{equation*}
 where
 \begin{equation*}
 NCT_{c} = N_c+Ran_{Dc}+NRan_{Dc}+IP_{nc}+P_{nc}+IRan_{Tc}+I_c
 \end{equation*}
 and
 \begin{equation*}
 NCT_{n} = N_n+Ran_{Tn}+Ran_{Dn}+NRan_{Dn}+IP_{nn}+P_{nn}+IRan_{Tn}+I_n
 \end{equation*}
 \item Fraction of Ran in the nucleus:
 \begin{equation*}
 \Phi_{Rann} = \frac{Ran_{Tn}+Ran_{Dn}+NRan_{Dn}+IRan_{Tn}}{Ran_{Dc}+NRan_{Dc}+IRan_{Tc}+Ran_{Tn}+Ran_{Dn}+NRan_{Dn}+IRan_{Tn}};
 \end{equation*}
 \item Fraction of cargo proteins in the nucleus:
 \begin{equation*}
 \Phi_{nn} = \frac{IP_{nn}+P_{nn}}{IP_{nc}+P_{nc}+IP_{nn}+P_{nn}}
 \end{equation*}
 \item Fraction of NTF2 in the nucleus:
 \begin{equation*}
 \Phi_{NTFn} = \frac{N_n+NRan_{Dn}}{N_c+NRan_{Dc}+N_n+NRan_{Dn}}
 \end{equation*}
 \item Fraction of importins in the nucleus:
 \begin{equation*}
 \Phi_{Impn} = \frac{IP_{nn}+IRan_{Tn}+I_n}{IP_{nc}+IRan_{Tc}+I_c+IP_{nn}+IRan_{Tn}+I_n}
 \end{equation*}
\end{enumerate}
\begin{figure}[h]
 \includegraphics[width=.9\textwidth]{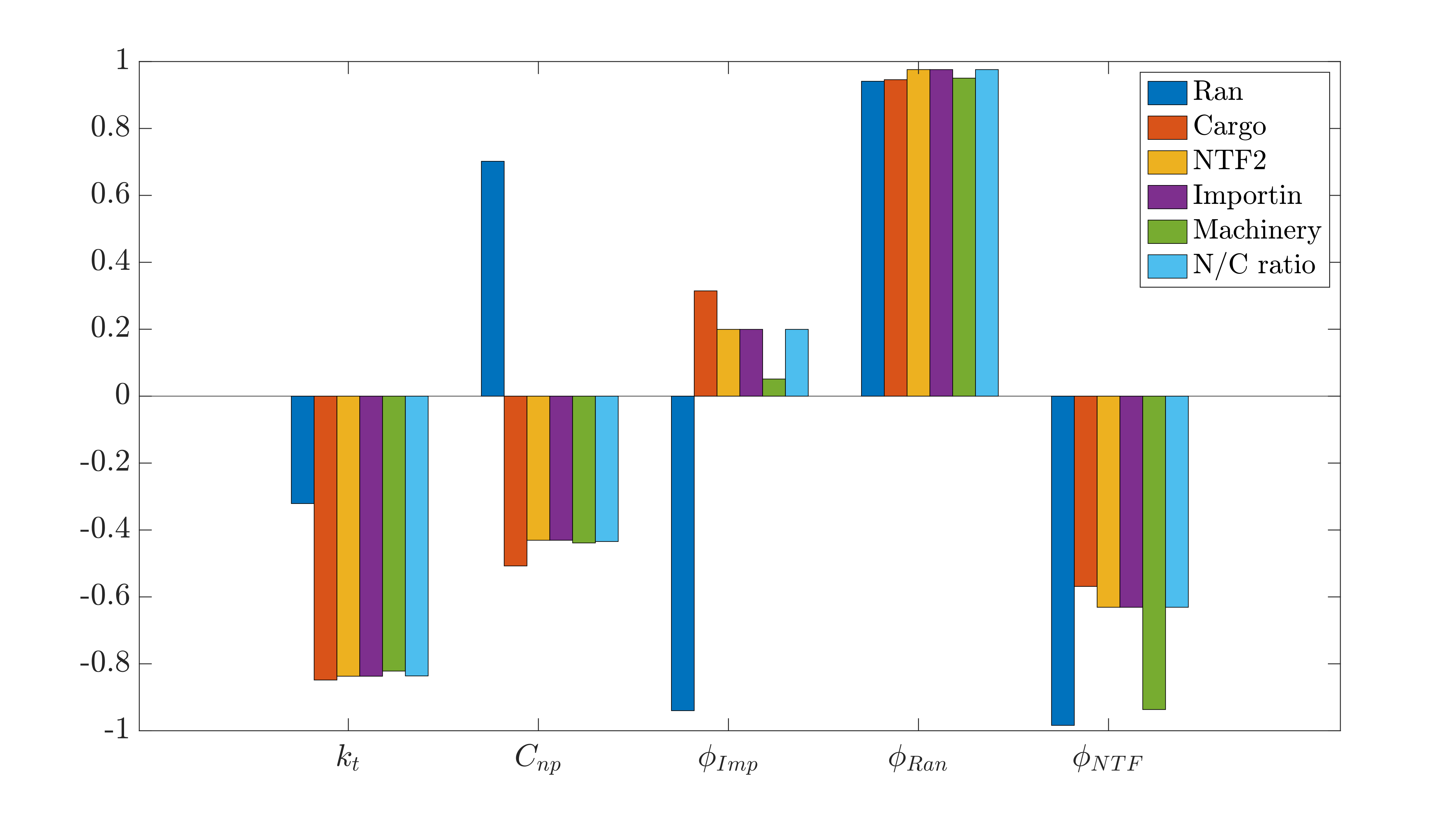}
 \caption{Sensitivity analysis of selected parameters of interest using the PRCC. QoIs are fractions of protein numbers in the nucleus. Ran: $\Phi_{Rann}$, Cargo: $\Phi_{nn}$, NTF2: $\Phi_{NTFn}$, Importin: $\Phi_{Impn}$, Machinery: $\Phi_{NCTn}$.}
 \label{Sensitivity1}
\end{figure}

\subsubsection{Gene fractions}
The correlation coefficients in Fig. \ref{Sensitivity} show that the gene fraction of importins, $\phi_{Imp}$, does not have a significant effect on the N/C ratio, indicating that at least within this range of parameters, the number of importins is unlikely to be a limiting factor for the nucleocytoplasmic transport; although Fig. \ref{Sensitivity1} shows that increasing $\phi_{Imp}$ significantly decreases the nuclear fraction of Ran. The gene fraction of NTF2, $\phi_{NTF}$, has a stronger effect on the N/C ratio, and the gene fraction of Ran proteins, $\phi_{Ran}$, has a powerful effect on the N/C ratio. Increasing $\phi_{NTF}$ decreases the N/C ratio, while increasing $\phi_{Ran}$ increases the N/C ratio.

Fig. \ref{Sensitivity1} shows the effects of increasing the gene fraction $\phi_{Ran}$ of Ran. Besides an increased N/C ratio, we also see an increase in the fraction $\Phi_{NCTn}$ of all nucleocytoplasmic transport machinery proteins in the nucleus. The increase of $\Phi_{NCTn}$ is the results of increase in the individual fractions $\Phi_{Rann}$, $\Phi_{nn}$, $\Phi_{NTFn}$, and $\Phi_{Impn}$, which means that increasing $\phi_{Ran}$ promotes nuclear localization of Ran, cargo proteins, NTF2, and importin.

Interestingly, we observed from simulation results that increasing $\phi_{Ran}$ increases the total nuclear concentration of Ran but decreases the total nuclear concentration of cargo proteins. Changes in the total nuclear concentration of NTF2 and importin are small and negligible. These observations suggest that the increase in the N/C ratio is mainly caused by the rise in the number of Ran proteins in the nucleus.

As we can see from Fig. \ref{Sensitivity1}, the overall effect of increasing $\phi_{NTF}$ is a decrease in the N/C ratio and the fraction $\Phi_{NCTn}$ of all nucleocytoplasmic transport machinery proteins in the nucleus. The decrease in $\Phi_{NCTn}$ is the result of the decrease in the individual fractions $\Phi_{Rann}$, $\Phi_{nn}$, $\Phi_{NTFn}$, and $\Phi_{Impn}$. 

Simulation results show that increasing the gene fraction $\phi_{NTF}$ of NTF2 leads to an increase in the nuclear concentration of free NTF2 and NTF2-RanGDP complex, which then increases the propensity of both directions in the reaction (\ref{R5}) that converts NTF2-RanGDP complex to free NTF2 and free RanGTP in the nucleus. The overall result is a decrease in the nuclear concentration of free RanGTP and the total nuclear concentration of Ran proteins. The decrease in the nuclear concentration of free RanGTP then decreases the unloading of free cargo proteins from importin-cargo complexes, which is shown by an increase in the nuclear concentration of importin-cargo complex and a decrease in the nuclear concentration of free cargo protein.

\subsubsection{Nuclear envelope permeability}
The correlation coefficients in Fig. \ref{Sensitivity} show that the nuclear envelope permeability coefficient $C_{np}$ also affects the N/C ratio. Increasing the nuclear envelope permeability decreases the N/C ratio within this range of parameters.

The nuclear envelope permeability coefficient $C_{np}$ directly affects the diffusion of proteins across the nuclear envelope (via reactions (\ref{R3}), (\ref{R6}), (\ref{R8}), (\ref{R9}), (\ref{R11})), which then affect the cytoplasmic and nuclear concentration of the corresponding protein species (importin-RanGTP complex, NTF2-RanGDP complex, importin, importin-cargo complex, NTF2). Simulation results indicate that increasing $C_{np}$ decreases the concentration differences across the nuclear envelope, which then affects multiple aspects in the nucleocytoplasmic transport. While we may perform further analysis to elucidate the exact mechanism through which $C_{np}$ affects the N/C ratio, such analysis is beyond the scope of this paper.

Fig. \ref{Sensitivity1} shows that, overall, increasing $C_{np}$ leads to an increase in the fraction of Ran in the nucleus, but leads to a decrease in the fraction of cargo in the nucleus. The net result is that the N/C ratio and the fraction of all nucleocytoplasmic transport machinery proteins in the nucleus decrease.

\subsubsection{Translation rate}
The correlation coefficients in Fig. \ref{Sensitivity} show that the translation rate $k_t$ also affects the N/C ratio and that within this range of parameters, increasing the translation rate decreases the N/C ratio. 

Fig. \ref{Sensitivity1} shows that increasing $k_t$ decreases the N/C ratio and the fraction $\Phi_{NCTn}$ of all nucleocytoplasmic transport machinery proteins in the nucleus. As $k_t$ increases, the number of nuclear cargo proteins increases more rapidly, and the number of proteins in the nucleocytoplasmic transport machinery also grows faster. However, the reactions of the nucleocytoplasmic transport process are limited by their reaction rates, so it becomes harder for these reactions to keep up with the faster-growing number of cargo proteins. Indeed, simulation results show that the concentration of free cargo proteins in the nucleus decreases as $k_t$ increases, whereas the concentration of free cargo proteins in the cytoplasm increases. Effectively, the nucleocytoplasmic transport machinery becomes saturated, so cargo proteins accumulate in the cytoplasm. The concentration of importin-cargo complexes in both the nucleus and the cytoplasm increases, but only slightly. Therefore, the overall result is that more cargo proteins become backed up in the cytoplasm, and the N/C ratio decreases.

\subsubsection{Reaction rates}
While we will not analyze in detail the effect of each of the reaction rates on the N/C ratio, we point out that, generally speaking, the reaction rates in the nucleus have a more substantial influence on the N/C ratio than the reaction rates in the cytoplasm, one noticeable exception being the reaction (\ref{R10}), which is the binding and unbinding of importin and cargo protein in the cytoplasm.

\subsection{The effect of NTF2 repression on nucleocytoplasmic transport and a compensation mechanism}\label{compensation}

Finally, we show an example comparing the simulation results from our nucleocytoplasmic transport and cell growth model to experimental observations, in which our model provides a possible explanation for these observed phenomena. In particular, we show that NTF2 is necessary for maintaining N/C ratio homeostasis and that increasing the expression of Ran may relieve the requirement for NTF2. These results are consistent with the experimental findings in \cite{paschal1997high}. Furthermore, we show that Ran's compensation effect requires the NTF2-independent permeability of RanGDP through the NPCs.

\subsubsection{Permeability of RanGDP through NPCs}

The original nucleocytoplasmic transport model from \cite{wang2017thermodynamic} does not assume the permeation of RanGDP directly through NPCs. Cytoplasmic RanGDP must first bind NTF2 to form a complex that can enter the nucleus. However, there are other possible models \cite{riddick2005systems} that would allow RanGDP to diffuse through NPCs without NTF2, albeit with much lower permeability. In the following simulations, we incorporate two additional reactions (\ref{R18}) and (\ref{R19}) into our model, shown here for the reader's convenience:
\begin{equation}\label{R18-1}
Ran_{Dc} \xrightleftharpoons [k_{18}^-]{k_{18}^+} Ran_{Dn},
\end{equation}
\begin{equation}\label{R19-1}
Ran_{Dn} \xrightleftharpoons [k_{19}^-]{k_{19}^+} Ran_{Tn}.
\end{equation}
The reaction (\ref{R18-1}) describes the diffusion of RanGDP in and out of the nucleus through NPCs, and (\ref{R19-1}) describes the reversible phosphorylation of RanGDP into RanGTP in the nucleus, mediated by RanGEF.

Note that in (\ref{R18-1}), the diffusion rate of RanGDP through NPCs is controlled by the coefficient $C_{Ran}$. Based on the permeability constants listed in \cite{riddick2005systems}, we set $C_{Ran}=3\times 10^{-2}$. In order to model the effects of no RanGDP diffusion through NPCs, we set $C_{Ran}=0$ as appropriate.

\subsubsection{NTF2 is necessary for maintaining N/C ratio homeostasis}
\begin{figure}[h]
 \includegraphics[width=.8\textwidth]{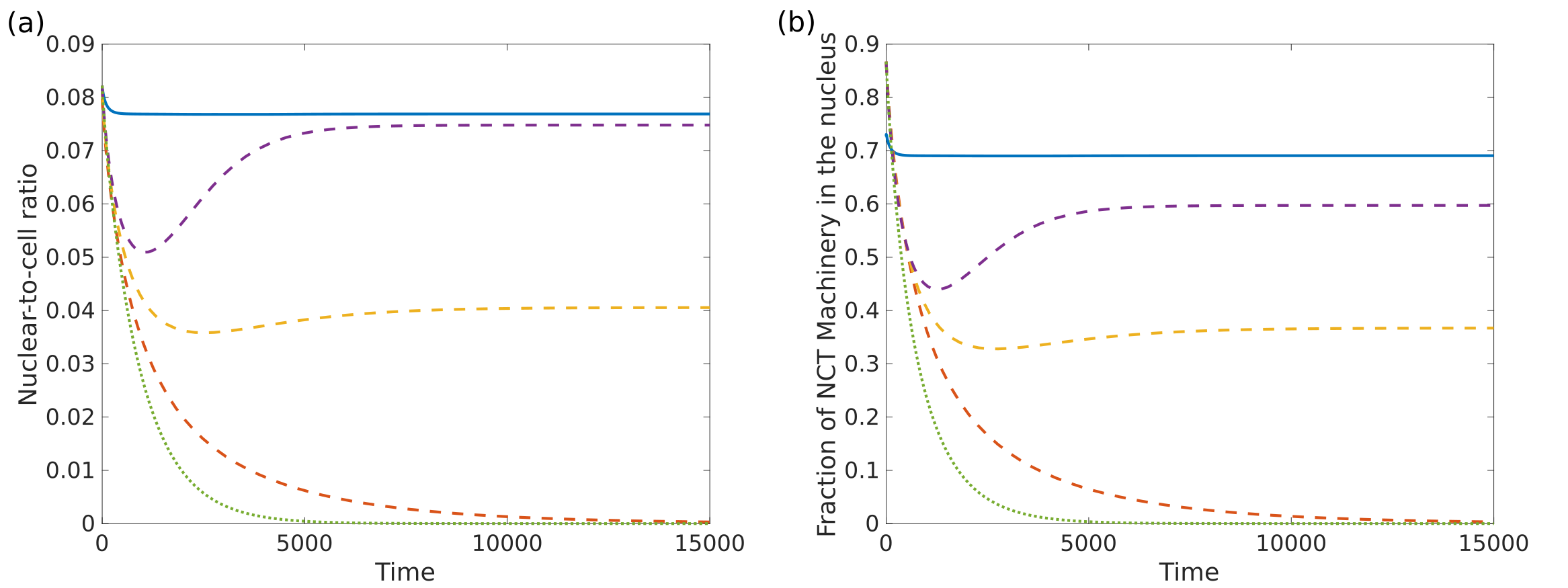}
 \caption{Effects of complete NTF2 repression. (a) N/C ratio; (b) Fraction of all nucleocytoplasmic transport machinery proteins in the nucleus. Dashed lines: $\phi_{NTF}=0$, $C_{Ran}=3\times 10^{-2}$, and $\phi_{Ran}=0.015$ (red), $0.029$ (yellow), and $0.044$ (purple). Dotted line: $\phi_{NTF}=0$, $C_{Ran}=0$ and $\phi_{Ran}=0.044$. Solid line: $\phi_{NTF}=0.02$, $C_{Ran}=3\times 10^{-2}$, and $\phi_{Ran}=0.015$ (normal values of the parameters, for comparison).}
 \label{NCComparisonRan}
\end{figure}

In \cite{paschal1997high}, the authors found that NTF2 is required for viability and that repression of NTF2 expression results in a substantial growth inhibition of cells. Our simulation results show that complete repression of NTF2 (by setting the gene fraction $\phi_{NTF}=0$ in the model) results in the N/C ratio eventually approaching zero (Fig. \ref{NCComparisonRan} (a), red dashed line), meaning that the nucleus will ultimately collapse; therefore, the cell cannot survive. The fraction of nucleocytoplasmic transport machinery proteins in the nucleus also approaches zero (Fig. \ref{NCComparisonRan} (b), red dashed line), indicating that the decrease of the nuclear volume is caused by a decline in the nucleocytoplasmic transport to zero. These results are consistent with experimental observations. In reality, repression of NTF2 expression results in distortion and apparent fragmentation of the nucleus (Fig. 3 in \cite{paschal1997high}).

\subsubsection{Compensation of NTF2 repression requires NPC permeability of RanGDP}

In \cite{paschal1997high}, the authors also found that high levels of Ran can relieve the need for NTF2 and rescue growth in cells with NTF2 repression. Our simulation results support this finding. Indeed, the dashed lines in Fig. \ref{NCComparisonRan} show that increasing the gene fraction $\phi_{Ran}$ of Ran can restore the N/C ratio of NTF2-repressed cells to approximately the normal N/C ratio, although achieving such compensation requires $\phi_{Ran}$ to be increased to three times its nominal value. However, as shown by the dotted line in Fig. \ref{NCComparisonRan}, the compensation effect of Ran overexpression depends on the NTF2-independent permeability of RanGDP through the NPCs. If we set $C_{Ran}=0$, which means that free RanGDP cannot pass through NPCs, then the N/C ratio will again approach zero.

\section{Discussion}
As stated in the Introduction, our model provides insights into several key aspects of nucleocytoplasmic transport, cell growth, and N/C ratio homeostasis. We have shown that in growing cells, the permeability of the nuclear envelope must increase to maintain a near-constant N/C ratio, that several parameters affecting the nucleocytoplasmic transport mechanism and gene translation also affect the N/C ratio, and that Ran may compensate for a lack of NTF2 in the nucleocytoplasmic transport mechanism only if RanGDP is allowed to translocate through the nuclear envelope independent of NTF2. 

In the following, we discuss the limitations of our current model that lead to future research directions.

In this model, ribosome biogenesis is modeled as a purely cytoplasmic process, whereas in reality, ribosomal proteins are imported into the nucleus, where they are combined with rRNAs to form ribosomal subunits; the subunits are then exported back to the cytoplasm to form mature ribosomes. (See, for example, \cite{chaker2018assembly,chaker2019assembly,woolford2013ribosome} for reviews of the ribosome biogenesis process.) This \textit{shuffling} of ribosomal proteins means ribosome biogenesis interacts with the nucleocytoplasmic transport mechanism and may affect the N/C ratio. The assembly of the ribosomal subunits in the nucleus and its effect on the N/C ratio have been modeled in previous research \cite{wu2022correlation}, in which the nucleocytoplasmic transport process is vastly simplified.

Similarly, in this model, the dynamics of the RanGAP- and RanGEF-mediated reactions (\ref{R2}) (\ref{R5}) are not modeled in detail. Therefore, our modeling of the two reactions may not represent a complete picture of the biochemical pathways. A more detailed model of the RanGEF-mediated reactions (\ref{R5}) can be found in \cite{gorlich2003characterization}.

In future research, we plan to expand this model by incorporating details of the ribosome biogenesis process, as well as details of the 
RanGAP and RanGEF-mediated reactions. However, one drawback of these more realistic models is the large number of parameters, which makes experimental determination of the parameter values difficult.

As shown in Section \ref{sensitivity}, many of the parameters in this model significantly affect the N/C ratio. We assume these parameters are constant throughout the cell growth process. In reality, however, the cell may dynamically regulate many of these parameters via feedback and/or feedforward pathways. For instance, in yeast cells, the rate of ribosome biogenesis is regulated by the TOR and PKA pathways and depends on signals such as nutrient level and stress \cite{guerra2022torc1}. This regulatory process may change the translation rate and affect the nucleocytoplasmic transport and the N/C ratio. An interesting avenue for future research is to expand this model to explore the effects of regulatory networks on the nucleocytoplasmic transport process and the N/C ratio.

In Section \ref{permeability}, we modeled the growth of the nuclear envelope permeability in a macroscopic manner. Although we investigated the dynamics of NPC number growth in light of the available experimental evidence to support specific models, we do not explicitly model the assembly of each NPC. In reality, NPC assembly is a complex, multi-step process \cite{dultz2008systematic,guglielmi2020nuclear,otsuka2018mechanisms,otsuka2023quantitative}. In particular, \textit{de novo} NPC assembly starts from \textit{inside} the nucleus, where specific protein complexes are recruited to the inner nuclear membrane to initiate the inside-out extrusion of the nuclear membranes \cite{otsuka2018mechanisms}. Therefore, similar to the case of ribosome biogenesis, the NPC assembly process also interacts with the nucleocytoplasmic transport mechanism and may affect the N/C ratio.

In summary, this study aims to understand the interaction between cellular physiological processes and the N/C ratio. Our work helps to explain the N/C ratio homeostasis and fluctuations within a broader perspective of cellular functions such as metabolism, transportation, growth, and the corresponding regulatory processes.

\section{Acknowledgments}
We acknowledge funding from NSF grant MCB-2213583.
\printbibliography
\appendix

\section{Complete list of reactions in the model}\label{reactionlist}

\subsection{Description of reactions in the model}
\subsubsection{Model component I: nucleocytoplasmic transport}
\begin{enumerate}
 \item Binding and unbinding of NTF2 ($N_c$) and RanGDP ($Ran_{Dc}$) in the cytoplasm;
 \item Reversible conversion of importin-RanGTP complex ($IRan_{Tc}$) into importin ($I_c$) and RanGDP ($Ran_{Dc}$) in the cytoplasm, mediated by RanGAP;
 \item Diffusion of the importin-RanGDP complex ($IRan_{Tn}$ and $IRan_{Tc}$) through NPCs;
 \item Binding and unbinding of importin ($I_n$) and RanGTP ($Ran_{Tn}$) in the nucleus;
 \item Reversible conversion of NTF2-RanGDP complex ($NRan_{Dn}$) into NTF2 ($N_n$) and RanGTP ($Ran_{Tn}$) in the nucleus, mediated by RanGEF;
 \item Diffusion of NTF2-RanGDP complex ($NRan_{Dc}$ and $NRan_{Dn}$) through NPCs;
 \item Binding and unbinding of importin ($I_n$) and nuclear cargo protein ($P_{nn}$) in the nucleus;
 \item Diffusion of importin ($I_c$ and $I_n$) through NPCs;
 \item Diffusion of importin-cargo complex ($IP_{nc}$ and $IP_{nn}$) through NPCs;
 \item Binding and unbinding of importin ($I_c$) and cargo protein ($P_{nc}$) in the cytoplasm;
 \item Diffusion of NTF2 ($N_n$ and $N_c$) through NPCs;
\end{enumerate}
\subsubsection{Model component II: cell growth}
\begin{enumerate}
\setcounter {enumi}{11}
 \item Autocatalytic biogenesis of ribosomes ($R_c$);
 \item Synthesis of nuclear cargo proteins ($P_{nc}$);
 \item Synthesis of cytoplasmic \textit{housekeeping} proteins ($P_c$);
 \item Synthesis of importins ($I_c$);
 \item Synthesis of Ran proteins ($Ran_{Dc}$);
 \item Synthesis of NTF2 proteins ($N_c$);
\end{enumerate}
\subsubsection{Model component III: additional reactions in nucleocytoplasmic transport}
\begin{enumerate}
\setcounter {enumi}{17}
\item Diffusion of RanGDP ($Ran_{Dc}$ and $Ran_{Dn}$) through NPCs;
\item Reversible conversion of RanGDP ($Ran_{Dn}$) into RanGTP ($Ran_{Tn}$) in the nucleus, mediated by RanGEF.
\end{enumerate}
\subsection{List of reactions}
\begingroup
\allowdisplaybreaks
\begin{align}
 \label{R1}
  N_c+Ran_{Dc} &\xrightleftharpoons [k_1^-]{k_1^+} NRan_{Dc} \\
 \label{R2}
  IRan_{Tc} &\xrightleftharpoons [k_2^-]{k_2^+} I_c+Ran_{Dc} \\
 \label{R3}
  IRan_{Tn} &\xrightleftharpoons [k_3^-]{k_3^+} IRan_{Tc} \\
 \label{R4}
  I_n+Ran_{Tn} &\xrightleftharpoons [k_4^-]{k_4^+} IRan_{Tn} \\
 \label{R5}
  NRan_{Dn} &\xrightleftharpoons [k_5^-]{k_5^+} N_n+Ran_{Tn} \\
 \label{R6}
  NRan_{Dc} &\xrightleftharpoons [k_6^-]{k_6^+} NRan_{Dn} \\
 \label{R7}
  IP_{nn} &\xrightleftharpoons [k_7^-]{k_7^+} I_n+P_{nn} \\
 \label{R8}
  I_c &\xrightleftharpoons [k_8^-]{k_8^+} I_n \\
 \label{R9}
  IP_{nc} &\xrightleftharpoons [k_9^-]{k_9^+} IP_{nn} \\
 \label{R10}
  I_c+P_{nc} &\xrightleftharpoons [k_{10}^-]{k_{10}^+} IP_{nc} \\
 \label{R11}
  N_n &\xrightleftharpoons [k_{11}^-]{k_{11}^+} N_c \\
 \label{R12}
  R_c &\xrightarrow{k_{12}} 2R_c \\
 \label{R13}
  R_c &\xrightarrow{k_{13}} R_c+P_{nc} \\
 \label{R14}
  R_c &\xrightarrow{k_{14}} R_c+P_c \\
 \label{R15}
  R_c &\xrightarrow{k_{15}} R_c+I_c \\
 \label{R16}
  R_c &\xrightarrow{k_{16}} R_c+Ran_{Dc} \\
 \label{R17}
  R_c &\xrightarrow{k_{17}} R_c+N_c \\
 \label{R18}
  Ran_{Dc} &\xrightleftharpoons [k_{18}^-]{k_{18}^+} Ran_{Dn} \\
 \label{R19}
  Ran_{Dn} &\xrightleftharpoons [k_{19}^-]{k_{19}^+} Ran_{Tn}
\end{align}
\endgroup
\subsection{Comparison to the original nucleocytoplasmic transport model}\label{modelcomparison}
In this section, we give a brief comparison explaining which reactions and species are new compared to the nucleocytoplasmic transport model detailed in \cite{wang2017thermodynamic}. 

Building on this previous nucleocytoplasmic transport model, our model incorporates the autocatalytic replication of ribosomes and expression of other proteins, as well as the diffusion and phosphorylation of free RanGDPs. Whereas the nucleocytoplasmic transport model of \cite{wang2017thermodynamic} consists of the biomolecule species $x_1\sim x_{14}$ described in Appendix \ref{vlist} and reactions (\ref{R1})$\sim$(\ref{R11}), our extended model augments this with the biomolecule species $x_{15}\sim x_{17}$ and reactions (\ref{R12})$\sim$(\ref{R19}). In particular, this yields a system of 17 biomolecule species and 19 equations in total.

\section{List of protein species}\label{vlist}

\begin{table}[h]\label{cspecies}
\begin{tabular}{ | l | l | l | l | l | l | l | l | l | l | } 
  \hline
  Species & $N_c$ & $Ran_{Dc}$ & $NRan_{Dc}$ & $IP_{nc}$ & $P_{nc}$ & $IRan_{Tc}$ & $I_c$ & $R_c$ & $P_c$ \\
  \hline
  Component & $x_1$ & $x_2$ & $x_3$ & $x_4$ & $x_5$ & $x_6$& $x_7$ & $x_{15}$ & $x_{16}$ \\
  \hline
\end{tabular}
\vspace{1ex}
\caption{List of cytoplasmic species.}
\end{table}

\begin{table}[h]\label{nspecies}
\begin{tabular}{ | l | l | l | l | l | l | l | l | l | } 
  \hline
  Species & $N_n$ & $Ran_{Tn}$ & $NRan_{Dn}$ & $IP_{nn}$ & $P_{nn}$ & $IRan_{Tn}$ & $I_n$ & $Ran_{Dn}$ \\
  \hline
  Component & $x_{8}$ & $x_{9}$ & $x_{10}$ & $x_{11}$ & $x_{12}$ & $x_{13}$ & $x_{14}$ & $x_{17}$ \\
  \hline
\end{tabular}
\vspace{1ex}
\caption{List of nuclear species.}
\end{table}

\section{List of propensity functions}\label{plist}

\begin{longtable}[c]{| c | c | c |}
\hline
Reaction & Forward propensity function & Reverse propensity function \\
\hline
\endfirsthead

\hline
Reaction & Forward propensity function & Reverse propensity function \\
\hline
\endhead

(\ref{R1}) & $\lambda_1^+(\mathbf{x})=\dfrac{k_1^+}{N_AV_{cyto}(\mathbf{x})}x_1x_2$ & $\lambda_1^-(\mathbf{x})=k_1^-x_3$ \\
\hline
(\ref{R2}) & $\lambda_2^+(\mathbf{x})=k_2^+x_6=k_{2,0}^+\theta^{-1}x_6$ & $\lambda_2^-(\mathbf{x})=\dfrac{k_2^-}{N_AV_{cyto}(\mathbf{x})}x_7x_2$ \\
\hline
(\ref{R3}) & $\lambda_3^+(\mathbf{x})=\dfrac{a(\mathbf{x})}{V_n(\mathbf{x})}x_{13}$ & $\lambda_3^-(\mathbf{x})=\dfrac{a(\mathbf{x})}{V_{cyto}(\mathbf{x})}x_6$ \\
\hline
(\ref{R4}) & $\lambda_4^+(\mathbf{x})=\dfrac{k_4^+}{N_AV_n(\mathbf{x})}x_{14}x_9$ & $\lambda_4^-(\mathbf{x})=k_4^-x_{13}$ \\
\hline
(\ref{R5}) & $\lambda_5^+(\mathbf{x})=k_5^+x_{10}=k_{5,0}^+\theta x_{10}$ & $\lambda_5^-(\mathbf{x})=\dfrac{k_5^-}{N_AV_n(\mathbf{x})}x_8x_9$ \\
\hline
(\ref{R6}) & $\lambda_6^+(\mathbf{x})=\dfrac{a(\mathbf{x})}{V_{cyto}(\mathbf{x})}x_3$ & $\lambda_6^-(\mathbf{x})=\dfrac{a(\mathbf{x})}{V_n(\mathbf{x})}x_{10}$ \\
\hline
(\ref{R7}) & $\lambda_7^+(\mathbf{x})=k_7^+x_{11}$ & $\lambda_7^-(\mathbf{x})=\dfrac{k_7^-}{N_AV_n(\mathbf{x})}x_{14}x_{12}$ \\
\hline
(\ref{R8}) & $\lambda_8^+(\mathbf{x})=\dfrac{a(\mathbf{x})}{V_{cyto}(\mathbf{x})}x_7$ & $\lambda_8^-(\mathbf{x})=\dfrac{a(\mathbf{x})}{V_n(\mathbf{x})}x_{14}$ \\
\hline
(\ref{R9}) & $\lambda_9^+(\mathbf{x})=\dfrac{a(\mathbf{x})}{V_{cyto}(\mathbf{x})}x_4$ & $\lambda_9^-(\mathbf{x})=\dfrac{a(\mathbf{x})}{V_n(\mathbf{x})}x_{11}$ \\
\hline
(\ref{R10}) & $\lambda_{10}^+(\mathbf{x})=\dfrac{k_{10}^+}{N_AV_{cyto}(\mathbf{x})}x_7x_5$ & $\lambda_{10}^-(\mathbf{x})=k_{10}^-x_4$ \\
\hline
(\ref{R11}) & $\lambda_{11}^+(\mathbf{x})=\dfrac{a(\mathbf{x})}{V_n(\mathbf{x})}x_8$ & $\lambda_{11}^-(\mathbf{x})=\dfrac{a(\mathbf{x})}{V_{cyto}(\mathbf{x})}x_1$ \\
\hline
(\ref{R12}) & $\lambda_{12}^+(\mathbf{x})=k_{12}^+x_{15}=k_t\phi_rx_{15}$ & - \\
\hline
(\ref{R13}) & $\lambda_{13}^+(\mathbf{x})=k_{13}^+x_{15}=k_t\phi_nx_{15}$ & - \\
\hline
(\ref{R14}) & $\lambda_{14}^+(\mathbf{x})=k_{14}^+x_{15}=k_t\phi_cx_{15}$ & - \\
\hline
(\ref{R15}) & $\lambda_{15}^+(\mathbf{x})=k_{15}^+x_{15}=k_t\phi_{Imp}x_{15}$ & - \\
\hline
(\ref{R16}) & $\lambda_{16}^+(\mathbf{x})=k_{16}^+x_{15}=k_t\phi_{Ran}x_{15}$ & - \\
\hline
(\ref{R17}) & $\lambda_{17}^+(\mathbf{x})=k_{17}^+x_{15}=k_t\phi_{NTF}x_{15}$ & - \\
\hline
(\ref{R18}) & $\lambda_{18}^+(\mathbf{x})=\dfrac{C_{Ran}a(\mathbf{x})}{V_{cyto}(\mathbf{x})}x_2$ & $\lambda_{18}^-(\mathbf{x})=\dfrac{C_{Ran}a(\mathbf{x})}{V_n(\mathbf{x})}x_{17}$ \\
\hline
(\ref{R19}) & $\lambda_{19}^+(\mathbf{x})=k_{19}^+x_{17}=k_{19,0}^+\theta x_{17}$ & $\lambda_{19}^-(\mathbf{x})=k_{19}^-x_9$ \\
\hline
\caption{List of propensity functions}
\label{propensitylist}
\end{longtable}

\section{List of parameter values}
\begin{table}[h]
\begin{tabular}{ | c | c | c | c |}
 \hline
 Parameter & Value & Parameter & Value \\
 \hline
 $k_1^+$ & $0.1$ nM$^{-1}$s$^{-1}$ & $k_1^-$ & $2.5$ s$^{-1}$ \\
 \hline
 $k_2^+$ & $300\theta^{-1}$ s$^{-1}$ & $k_2^-$ & $1$ nM$^{-1}$s$^{-1}$ \\
 \hline
 $k_4^+$ & $0.1$ nM$^{-1}$s$^{-1}$ & $k_4^-$ & $0.1$ s$^{-1}$ \\
 \hline
 $k_5^+$ & $200\theta$ s$^{-1}$ & $k_5^-$ & $1$ nM$^{-1}$s$^{-1}$ \\
 \hline
 $k_7^+$ & $20$ s$^{-1}$ & $k_7^-$ & $1$ nM$^{-1}$s$^{-1}$ \\
 \hline
 $k_{10}^+$ & $1$ nM$^{-1}$s$^{-1}$ & $k_{10}^-$ & $20$ s$^{-1}$ \\
 \hline
 $k_{19}^+$ & $100\theta$ s$^{-1}$ & $k_{19}^-$ & $1$ s$^{-1}$ \\
 \hline
\end{tabular}
\vspace{1ex}
\caption{List of reaction rates.}
\end{table}

\begin{table}[h]
\begin{tabular}{ | c | c | c | c | c | c |}
 \hline
 Parameter & Value & Parameter & Value & Parameter & Value \\
 \hline
 $\phi_r$ & $0.2$ & $\phi_n$ & $0.0804$ & $\phi_c$ & $0.6777$ \\
 \hline
 $\phi_{Imp}$ & $0.0078$ & $\phi_{Ran}$ & $0.0146$ & $\phi_{NTF}$ & $0.0195$ \\
 \hline
\end{tabular}
\vspace{1ex}
\caption{List of gene fractions.}
\end{table}

\end{document}
